\documentclass[epj]{svjour}
\usepackage{graphicx,epsfig}
\usepackage{amsmath,amssymb}
\usepackage{arydshln}
\usepackage{multirow}

\newcommand{\ord}[1]{\mathcal{O}\left({#1}\right)}
\newcommand{\BR}{{\cal B}}
\newcommand{\no}{\nonumber}

\begin{document}

\title{Split-Family SUSY, $U(2)^5$ Flavour Symmetry and Neutrino Physics}

\author{
Joel~Jones-P\'erez\inst{1,2}}

\institute{
Departament de F\'{\i}sica Te\`orica and IFIC, Universitat de Val\`encia-CSIC, E-46100, Burjassot, Spain
\and
CERN, Theory Division, 1211 Geneva 23, Switzerland
}

\mail{joel.jones@uv.es}

\date{Received: date / Revised version: date}

\abstract{
In split-family SUSY, one can use a $U(2)^3$ symmetry to protect flavour observables in the quark sector from SUSY contributions. However, attempts to extend this procedure to the lepton sector by using an analogous $U(2)^5$ symmetry fail to reproduce the neutrino data without introducing some form of fine-tuning. In this work, we solve this problem by shifting the $U(2)^2$ symmetry acting on leptons towards the second and third generations. This allows neutrino data to be reproduced without much difficulties, as well as protecting the leptonic flavour observables from SUSY. Key signatures are a $\mu\to e\gamma$ branching ratio possibly observable in the near future, as well as having selectrons as the lightest sleptons.
}

\maketitle

\section{Introduction}

The first run of the LHC has already been completed, with no signs of any kind of new physics. Although this puts significant pressure on most types of physics beyond the Standard Model, models based on low-energy Supersymmetry (SUSY) seem to be at the heart of most discussions. The reason for this is the apparent difficulty of the latter to completely remove fine-tuning in the Higgs sector~\cite{manycites2}. 

The question of what amount of tuning is acceptable, if any at all, is still unresolved. However, it is generally acknowledged that to minimise the problem one requires light stops, light higgsinos and not too heavy gluinos~\cite{Papucci:2011wy} (see~\cite{Antusch:2012gv,Feng:1999mn} for alternative approaches). This, combined with current LHC bounds on SUSY masses~\cite{TheATLAScollaboration:2013fha,CMS:2013gea}, suggests a departure from simple SUSY models. In particular, if first generation squark masses are pushed to higher values, it might be necessary to introduce a split-family spectrum for sfermions~\cite{manycites1}.

Demanding such a spectrum raises a new issue regarding flavour. Generally, in order to suppress SUSY contributions to flavoured processes, the Minimal Flavour Violation (MFV) ansatz is invoked~\cite{D'Ambrosio:2002ex}. This, however, might not be acceptable any more in a split-family scenario, as MFV naturally demands all sfermions to have equal masses, with family splitting appearing at first order in the spurion expansion. In order to obtain non-degenerate masses, one requires a cancellation between the the leading two contributions to the third generation soft masses\footnote{Of course, RGE effects also introduce further splitting, up to a certain degree.}. This suggests that split-family SUSY cannot be combined with MFV without recurring to some amount of fine-tuning in the flavour sector.

An alternative ansatz without this problem was found in~\cite{Barbieri:2011ci}, for quarks. Here, instead of the $U(3)^3$ flavour symmetry of MFV, a $U(2)^3$ symmetry was imposed, acting on the first two generations. This would allow the separatation of the (light) soft mass of the third generation squarks from that of the (heavy) first two, simultaneously suppressing SUSY flavour effects to a sufficient level. This work was followed by further phenomenological studies in the quark sector~\cite{Barbieri:2011fc,Barbieri:2012uh,Buras:2012sd}, as well as studies of perturbations due to RGE effects~\cite{Blankenburg:2012ah}.

To have a full understanding of this ansatz, it is necessary to apply it also on the lepton sector. It was shown in~\cite{Blankenburg:2012nx} that this is not straightforward, leading to a somewhat different approach. Here, the starting point was set to be $U(3)^5$ (i.e.\, MFV), with a breaking of the symmetry to $U(2)^5$ on the Yukawa sector, and to $O(3)_L$ on the neutrino sector. Although this approach allowed the construction of the same Yukawa matrices as in $U(2)^5$, along with a satisfactory reproduction of the neutrino oscillation parameters, it was found that the soft terms suffered the same problem as MFV: in order to obtain a split-family sfermion spectrum, one requires a cancellation.

In this work, we build an alternative framework to that of~\cite{Blankenburg:2012nx}, such that no cancellation is needed to split the sfermion generations, neutrino data is easily reproduced, and flavour is still protected from large SUSY contributions. The method is based on shifting the $U(2)^2$ symmetries acting on leptons towards the second and third generations.

To understand our motivation, let us briefly review what we know of flavour in the leptonic sector. The mixing of leptons is observed in neutrino oscillation experiments, which have measured three mixing angles~\cite{GonzalezGarcia:2012sz}:
\begin{align}
 s^2_{12} = 0.306^{+0.012}_{-0.012}~, & & s^2_{13} = 0.0229^{+0.0020}_{-0.0019}~, \nonumber
 \end{align}
 \begin{equation}
 s^2_{23} = 0.446^{+0.007}_{-0.007}\,\oplus\, 0.587^{+0.032}_{-0.037}~.
\end{equation}
These are very different from those of quarks, with the smallest lepton mixing angle having a size similar to the largest one for quarks.

Neutrino experiments have also measured two mass squared differences, although the sign of one of them has yet to be determined. In the so-called normal hierarchy, the mass differences are:
\begin{eqnarray}
 \Delta m^2_{\rm atm} &=& (2.417^{+0.013}_{-0.013})\times10^{-3}~{\rm eV}^2, \nonumber \\
 \Delta m^2_{\rm sol} &=& (7.45^{+0.19}_{-0.16})\times10^{-5}~{\rm eV}^2~,
\end{eqnarray}
from which we can extract a small parameter:
\begin{equation}
\zeta^2=\Delta m^2_{\rm sol}/\Delta m^2_{\rm atm}~. 
\end{equation}

On the basis where the charged lepton Yukawa is diagonal, neutrino masses and mixing can be extracted from the Majorana neutrino mass matrix $m_\nu$, appearing in the dimension-five effective operator. Still in the normal hierarchy, on the limit where $\zeta^2\to0$ and $s^2_{13}\to0$, we find:
\begin{multline}\label{eq-start.nh}
(m_\nu^\dagger m_\nu)_{\rm n.h.}\rightarrow m_{\rm light}^2 \cdot \left(\begin{array}{ccc}
1 & 0 & 0\\0 & 1 & 0\\0 & 0 & 1 
\end{array}\right)\\ + \Delta m^2_{\rm atm} \left(
\begin{array}{ccc}
0 & 0 & 0\\0 & s_{23}^2 & s_{23}c_{23}\\0 & s_{23}c_{23} & c_{23}^2 
\end{array}\right)~,
\end{multline}
where $m_{\rm light}$ is the lightest neutrino mass.

This equation gives us a very important hint on how to treat neutrinos. On one hand, for large $m_{\rm light}$, we find data prefers degenerate neutrino masses, and thus disfavours the application of any $U(2)$ symmetry. This motivated the approach used in~\cite{Blankenburg:2012nx}, where $U(3)_L$ was broken to $O(3)_L$ for neutrinos only. 

However, for vanishing $m_{\rm light}$, we can see that neutrinos require a connection between the second and third generations, not between the first and second. It is this fact that motivates this work.

In the following, we build a framework based on $U(2)^5$ flavour symmetries, in such a way that it acts on the first two generations in the quark sector, and on the last two in the lepton sector. The setup is described in Section~\ref{sec:quarks}, where we also show how this is applied on the quark sector. The lepton and slepton sectors are addressed in Sections~\ref{sec:lepton} and~\ref{sec:slepton}, respectively. Also in Section~\ref{sec:slepton}, we study lepton flavour violation (LFV) processes, and show that flavoured observables are indeed protected from large SUSY contributions. The neutrino sector is described in more detail in Appendices~\ref{app:nh} and~\ref{app:ih}. We conclude in Section~\ref{sec:conc}.

\section{Setup and Quark Sector}
\label{sec:quarks}

\begin{table*}
\begin{center}
\begin{tabular}{|c||c|c||c|c||c|c||c|c||c|c|}
\hline
& $\phi_Q$ & $\rho_Q$ & $\phi_u$ & $\rho_u$ & $\phi_d$ & $\rho_d$ & $\phi_L$ & $\rho_L$ & $\phi_e$ & $\rho_e$  \\
\hline 
$SU(3)_C$ & {\bf 3} & {\bf 3} & $\boldsymbol{\bar 3}$ & $\boldsymbol{\bar 3}$ & $\boldsymbol{\bar 3}$ & $\boldsymbol{\bar 3}$ &
1 & 1 & 1 & 1 \\
$SU(2)_L$ & {\bf 2} & {\bf 2} & 1 & 1 & 1 & 1 & {\bf 2} & {\bf 2} & 1 & 1 \\
$U(1)_Y$ & 1/6 & 1/6 & -2/3 & -2/3 & 1/3 & 1/3 & -1/2 & -1/2 & 1 & 1 \\
\hline \hline
$U(2)_Q$ & {\bf 2} & 1 & 1 & 1 & 1 & 1 & 1 & 1 & 1 & 1 \\
$U(2)_u$ & 1 & 1 & {\bf 2} & 1 & 1 & 1 & 1 & 1 & 1 & 1 \\
$U(2)_d$ & 1 & 1 & 1 & 1 & {\bf 2} & 1 & 1 & 1 & 1 & 1 \\
$U(2)_L$ & 1 & 1 & 1 & 1 & 1 & 1 & {\bf 2} & 1 & 1 & 1 \\
$U(2)_e$ & 1 & 1 & 1 & 1 & 1 & 1 & 1 & 1 & {\bf 2} & 1 \\
\hline
$U(1)_d$ & 0 & 0 & 0 & 0 & 0 & $\beta_1$ & 0 & 0 & 0 & 0 \\
$U(1)_L$ & 0 & 0 & 0 & 0 & 0 & 0 & 0 & $\beta_2$ & 0 & 0 \\
$U(1)_e$ & 0 & 0 & 0 & 0 & 0 & 0 & 0 & 0 & 0 & $\beta_3$ \\
\hline
\end{tabular}
\end{center}
\caption{Symmetries and SM superparticle content. For $U(1)$ symmetries, we show the actual charge of the field.}
\label{tab:particles}
\end{table*}
The flavour symmetry acting on the MSSM Lagrangian shall be $U(2)^5\otimes U(1)_d\otimes U(1)_L\otimes U(1)_e$. Apart from the Higgs doublets, $H_u$ and $H_d$, which do not transform under any flavour symmetry, we consider the matter content shown in Table~\ref{tab:particles}. These correspond to the MSSM quark and lepton superfields, with the $\phi_f$ transforming as doublets under some $U(2)_f$ symmetry. The $\rho_f$ are $U(2)^5$ singlets, some of them transforming under one of the three $U(1)_f$.

In the limit of exact symmetry, the only term with quark superfields allowed in the Superpotential is
\begin{equation}
 W=y_t\,\rho_Q\,\rho_u\,H_u,
\end{equation}
where $y_t$ is an $\ord{1}$ parameter. Thus, the superfield $\rho_Q$ corresponds to the third generation $Q_L$, and $\rho_u$ corresponds to the third generation $u^c_R$.

Reproducing the Yukawa couplings of the quark sector is straightforward. We proceed by adding spurions with the following transformation properties under the flavour symmetries:
\begin{align}
&\Delta Y_u \rightarrow (2,2,1,1,1)_{(0,0,0)}~, & 
 \Delta Y_d \rightarrow (2,1,2,1,1)_{(0,0,0)}~, \nonumber \\
&V_q \rightarrow (2,1,1,1,1)_{(0,0,0)}~,  & y_b \rightarrow (1,1,1,1,1)_{(-\beta_1,0,0)}~, &
\end{align}
where the $(a,b,c)$ subindices refer to the charges under $U(1)_d$, $U(1)_L$ and $U(1)_e$, respectively. These lead to the following terms in the Superpotential, invariant under the flavour symmetry:
\begin{multline}
 \delta W =  \left(\phi_Q\,V_q\,\rho_u+\phi_Q\,\Delta Y_u\,\phi_u\right)H_u \\
 +\left(y_b\,\rho_Q\,\rho_d +y_b\,\phi_Q\,V_q\,\rho_d+ \phi_Q\,\Delta Y_d\,\phi_d\right)H_d~,
\end{multline}
where we have neglected $\ord{1}$ constants. The spurions thus give rise to the Yukawa couplings of~\cite{Barbieri:2011ci}, namely:
\begin{align}
 Y_u=\left(\begin{array}{c:c}
 \Delta Y_u & x_t\,V_q \\\hdashline
 0 & y_t
\end{array}\right), & &
Y_d=\left(\begin{array}{c:c}
 \Delta Y_d & x_b\,V_q\,y_b \\\hdashline
 0 & y_b
\end{array}\right),
\end{align}
where $x_f$ are complex $\ord{1}$ constants that cannot be removed by spurion redefinition. Here, everything to the left of the vertical line has two columns, and everything above the horizontal line has two rows. One then identifies $\phi_Q$, $\phi_u$ and $\phi_d$ with the first two families of $Q_L$, $u_R^c$ and $d_R^c$. Notice that the $y_b$ spurion plays the role of providing a justification for the difference between the top and bottom quark masses.

Once we have the Yukawas built in this manner, the analysis and conclusions are analogous to those of the original $U(2)^3$ framework, for both the quark and squark sectors~\cite{Barbieri:2011ci,Barbieri:2011fc,Barbieri:2012uh,Buras:2012sd}. In particular, we can keep the parametrisation:
\begin{align}
 \Delta Y_f = \left(\begin{array}{cc}
c_f & s_f\,e^{i\alpha_f} \\
-s_f\,e^{-i\alpha_f} & c_f
\end{array}\right)\Delta Y_f^d~, & &
V_q = \left(\begin{array}{c}
 0 \\ 1\end{array}\right)\epsilon~,
\end{align}
where $\epsilon\sim\lambda_{\rm CKM}^2$, and $\Delta Y_f^d$ means that the matrix has been diagonalised. If we choose $s_u\lesssim s_d\sim\lambda_{\rm CKM}$, we reproduce the CKM mixing matrix.

\section{Lepton Sector}
\label{sec:lepton}

For the lepton sector, we introduce spurions with transformation properties analogous to those for quarks:
\begin{align}
\Delta Y_e \rightarrow (1,1,1,2,2)_{(0,0,0)}~, & & V_e \rightarrow (1,1,1,2,1)_{(0,0,0)}~, \nonumber \\
\lambda_L \rightarrow (1,1,1,1,1)_{(0,-\beta_2,0)}~, & &
 \lambda_e \rightarrow (1,1,1,1,1)_{(0,0,-\beta_3)}~,
\end{align}
from which we build the charged lepton Yukawa matrix.

The key assumption used to shift the $U(2)_f$ symmetries to the second and third generations lies on taking $(\lambda_L\lambda_e)=y_e$, i.e.\ the first generation Yukawa coupling. This gives the charged lepton Yukawa a different structure. When ordering the states in terms of ascending eigenvalues, and re-defining the spurions in order to absorb the $\ord{1}$ parameters, we find:
\begin{equation}
 Y_e=\left(\begin{array}{c::c}
 y_{e} & 0 \\\hdashline\hdashline
 V_e\,\lambda_e & \Delta Y_e
\end{array}\right)~.
\end{equation}
In this case, everything to the right of the double vertical lines has two columns, and everything under the double horizontal lines has two rows. Thus, the small value of the $\lambda_L\,\lambda_e$ spurion combination shifts the $U(2)_L\otimes U(2)_e$ symmetry in the lepton sector towards the second and third generations.

For the neutrino sector, we build the effective Majorana mass matrix using the same spurions as for the Yukawas. We find it necessary to introduce an additional spurion:
\begin{equation}
 \Delta_L \rightarrow (1,1,1,3,1)_{(0,0,0)}~, 
\end{equation}
such that the matrix in the dimension-five operator becomes:
\begin{equation}
 m_\nu=\left(\begin{array}{c::c}
 r_1\,\lambda_L^2 & r_2\,\lambda_L\,V^T_e \\\hdashline\hdashline
 r_2\,\lambda_L\,V_e & \Delta_L+V_e\,V_e^T
\end{array}\right)\kappa_\nu~.
\end{equation}
with $r_1$ and $r_2$ being the only $\ord{1}$ complex couplings that cannot be re-absorbed into the spurions. The spurion $\kappa_\nu$ acts as a complex parameter with mass dimension, breaking lepton number.

For the lepton sector, we find it more useful to use a different parametrisation from that for quarks. We keep $\Delta Y_e$ fully diagonal from the start, such that $V_e$ is  unaligned:
\begin{align}
 \Delta Y_e = \left(\begin{array}{cc}
 y_\mu & 0 \\
 0 & y_\tau
\end{array}
\right)~, & &
V_e =
\left(\begin{array}{c}
 s_e \\
 c_e
\end{array}
\right)\epsilon~.
\end{align}
Here we take the suppression in $V_e$ equal to that in $V_q$, with $s_e$ ($c_e$) being the sine (cosine) of an undetermined mixing angle $\theta_e$. As there is only one leptonic Yukawa, we find it possible to remove all phases from $Y_e$. The resulting structures are:
\begin{eqnarray}
\label{yuktexture} 
Y_e &=& \left(\begin{array}{ccc}
 y_e & 0 & 0 \\
 s_e\,\epsilon\,\lambda_e & y_\mu & 0 \\
 c_e\,\epsilon\,\lambda_e & 0 & y_\tau
\end{array}\right)~, \\
\label{majomat} 
m_\nu&=&\left(\begin{array}{ccc}
r_1\lambda_L^2 & r_2 s_e\lambda_L\epsilon & r_2 c_e\lambda_L\epsilon \\
r_2 s_e\lambda_L\epsilon & \multicolumn{2}{c}{\multirow{2}{*}{$U^T_\phi\,D\, U_\phi$}} \\
r_2 c_e\lambda_L\epsilon & & 
\end{array}\right)\kappa_\nu~,
\end{eqnarray}
where we have parametrised the $\Delta_L+V_eV_e^T$ combination using:
\begin{eqnarray}
D &=& \left(\begin{array}{cc}
 d_1\,e^{i\omega_1} & 0 \\ 0 & d_2\,e^{i\omega_2}
\end{array}\right)\epsilon^2~, \\
U_\phi &=& \left(\begin{array}{cc}
 c_\phi & s_\phi\,e^{i\omega_3} \\
 -s_\phi\,e^{-i\omega_3} & c_\phi
\end{array}\right)~.
\end{eqnarray}
The $d_i$, $s_\phi$ and $\omega_i$ parameters are functions of $s_e$ and the elements of $\Delta_L$. We fix our parametrisation such that $d_1<d_2$. Given the suppression of $V_eV_e^T$, we expect $d_2$ to be at least of $\ord{1}$. 

The study of the parameter space is non-trivial, and is carried out in more detail on Appendices~\ref{app:nh} and~\ref{app:ih}. We find it possible to reproduce the neutrino parameters for both normal and inverted hierarchies. However, the latter requires a strong connection between $\lambda_L$ and $d_2$, which are not related by any symmetry. Thus, we conclude that this framework favours the normal hierarchy, and shall present results valid only in this scenario.

As in leptonic MFV~\cite{Cirigliano:2005ck}, the charged lepton and neutrino data are used to obtain information about the spurion structure and suppression, to be used later for estimating new physics contributions to flavoured processes. To this end, the following parameters are relevant:
\begin{eqnarray}
 \lambda_L &=& 10^{-2} \\
 \lambda_e &=& y_e\,10^2 \\
  \Delta_L &=& \ord{\epsilon^2} \\
 s_e & = & (\rm Unconstrained)
\end{eqnarray}
Although the neutrino data does not put any bounds on $s_e$, a comparison with the quark sector suggests that taking $s_e\sim\ord{10^{-1}}$ is reasonable.

The choice of parameters that best describes neutrino data also has implications on the lightest neutrino mass. We find an upper bound $m_{\rm light}\lesssim10^{-2}$~eV, meaning that we expect no observation of neutrinoless double beta decay~\cite{Schwingenheuer:2012zs}, as well as no constraints from cosmology~\cite{Jimenez:2010ev}.

\section{Slepton Sector}
\label{sec:slepton}

Following the same philosophy as MFV, the spurions in the Superpotential also appear in the soft mass matrices and trilinear terms. This leads to the following structure for the soft mass matrices, at leading order and neglecting $\ord{1}$ couplings:
\begin{eqnarray}
 \frac{m^2_{\tilde L}}{m_h^2} &=& \left(\begin{array}{c::c}
 m_l^2/m_h^2 & \lambda^*_L\,V_e^T \\\hdashline\hdashline
\lambda_L\,V_e^* & I + \Delta Y_e^*\Delta Y_e^T + V_e^*V_e^T+\Delta_L^*\Delta_L
\end{array}\right)\delta_{\rm s} \\
\frac{m^2_{\tilde e_R^c}}{m_h^2} &=& \left(\begin{array}{c::c}
 m_l^2/m_h^2 & \lambda^*_e\,V_e^\dagger\Delta Y_e \\\hdashline\hdashline
\lambda_e\,\Delta Y_e^\dagger\,V_e & I + \Delta Y_e^\dagger\Delta Y_e
\end{array}\right)\delta_{\rm s}
\end{eqnarray}
Here, $m_l^2$ and $m_h^2$ are SUSY-breaking terms. As usual, we assume that the SUSY-breaking term associated to the $U(2)^5$ doublets is much heavier than the one for singlets (i.e.\ $m^2_h\gg m_l^2$). This can be achieved in principle through specific SUSY-breaking 
mechanisms, such as~\cite{Delgado:2011kr,Dudas:2013pja}. Notice that, in contrast to the squark sector, where the stops and sbottoms are expected to be light, here it is the selectron, and not the stau, the one which becomes the lightest slepton\footnote{For concreteness, we neglect possible RGE effects.}. 

The parameter $\delta_{\rm s}$ contains all possible spurion combination that can form singlets, such as:
\begin{equation}
 \delta_{\rm s} = 1+\sum_n c_n (\lambda_L\lambda_L^*)^n +\sum_n c'_n (\lambda_e\lambda_e^*)^n+\ldots~.
\end{equation}
For simplicity, we shall absorb $\delta_{\rm s}\approx 1$ into $m_h^2$.

In the following, we neglect the $\Delta_L^*\Delta_L$ contribution, as we expect it to be of order $\epsilon^4$. We also expect the size of $LR$-mixing effects to be $\ord{m_e/m_h}$, such that the left and right sectors do not mix. This means we can treat separately the $LL$ and $RR$ mixing. 

For $LL$ mixing, we neglect the contribution coming from the diagonalisation of $Y_e\,Y_e^\dagger$. As mentioned in Appendix~\ref{app:nh}, for low enough values of $\lambda_e$, the matrices that diagonalise $Y_e\,Y_e^\dagger$ are nearly diagonal. Thus, if we have $\mathcal{R}^{L\dagger} m^2_{\tilde L} \mathcal{R}^L = (m^2_{\tilde L})^{\rm diag}$, the mixing matrix has the form:
\begin{equation}
\label{sneutrinomix}
 \mathcal{R}^L\sim\left(\begin{array}{ccc}
1 & s_e\lambda_L\epsilon & c_e\lambda_L\epsilon \\
-s_e\lambda_L\epsilon & 1 & s_ec_e\frac{\epsilon^2}{y_\tau^2} \\
-c_e\lambda_L\epsilon & -s_ec_e\frac{\epsilon^2}{y_\tau^2} & 1
\end{array}\right)~,
\end{equation}
where we have neglected some real $\ord{1}$ couplings. Notice that due to the near degeneracy of the smuon and stau, the sleptonic $2-3$ mixing is not small. However, this should not be a problem for LFV processes, as the latter are expected to be very heavy.

For $RR$ mixing, notice the off-diagonal terms of $m^2_{\tilde e^c_R}$ are highly suppressed. However, in contrast to the $LL$ sector, the diagonalisation of $Y_e^\dagger\,Y_e$ does give a non-negligible contribution to the mixing. Taking these effects into account, we find:
\begin{equation}
\label{RRmix}
 \mathcal{R}^R\sim\left(\begin{array}{ccc}
1 & -s_e\lambda_e\epsilon/y_\mu & -c_e\lambda_e\epsilon/y_\tau \\
s_e\lambda_e\epsilon/y_\mu & 1 & 0 \\
c_e\lambda_e\epsilon/y_\tau & 0 & 1
\end{array}\right)~.
\end{equation}
For the $1-2$ sector, taking into account that $y_e/y_\mu\sim\ord{0.1\,\epsilon}$, we find that, contrary to leptonic MFV, here $RR$ mixing is about 50 times larger than $LL$ mixing. For the $1-3$ sector, both types of mixing are comparable.


\subsection{Lepton Flavour Violation}

We consider two types of physical processes where slepton flavour mixing plays a role: high-energy and low-energy LFV.

For high-energy LFV, we are interested in selectron decay into a neutralino plus a muon or tau lepton. These are possible both at the LHC and linear $e^+e^-$ colliders (for recent studies, see~\cite{Abada:2010kj,Abada:2012re}).

\begin{figure*}[tbp]
\begin{center}
\includegraphics[width=0.45\textwidth]{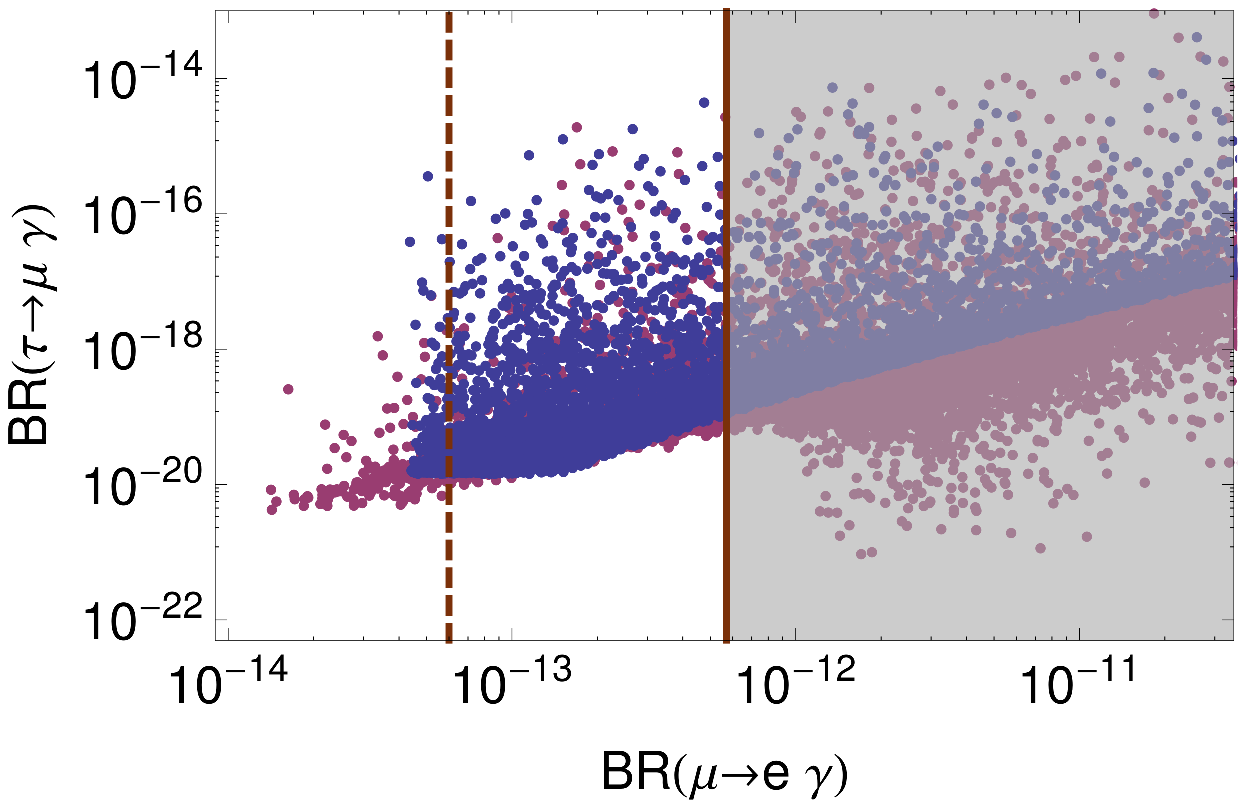} \quad
\includegraphics[width=0.45\textwidth]{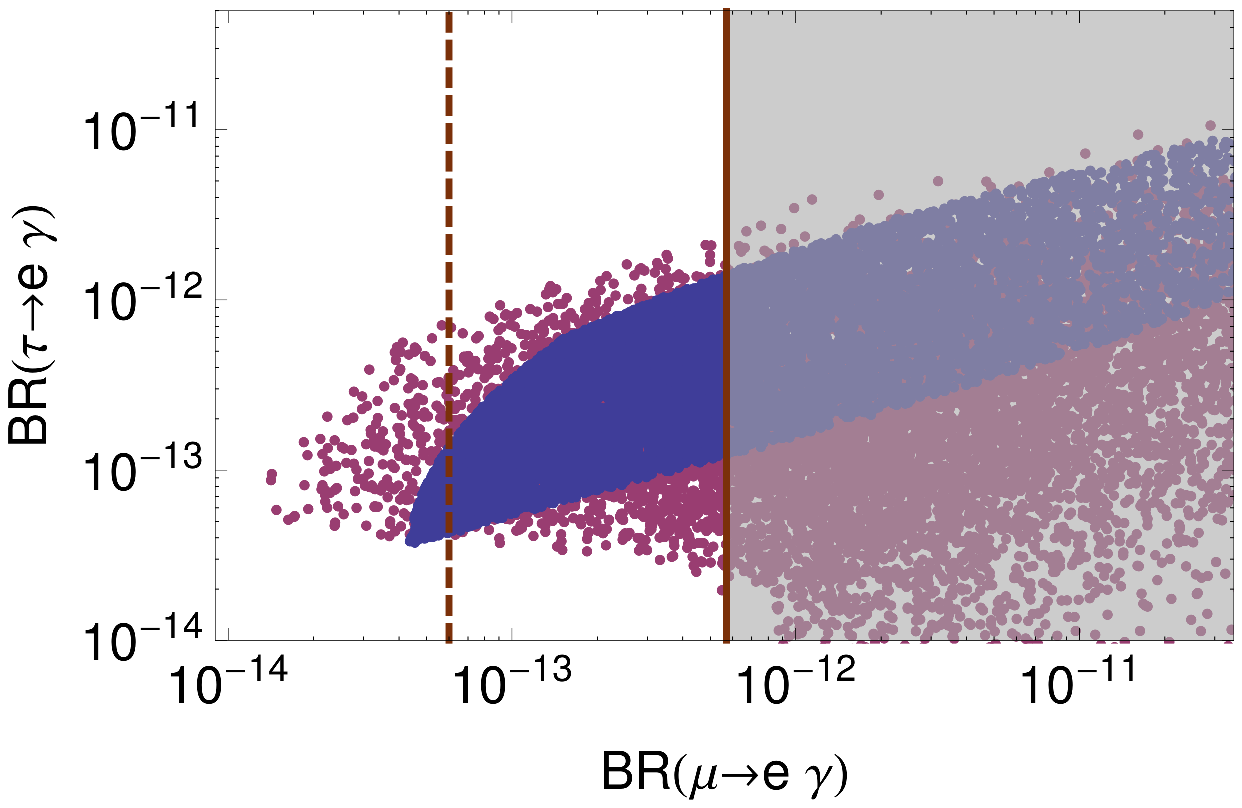}
\end{center}
\caption{Branching ratios for radiative LFV. Left: $\tau\to\mu\gamma$ vs $\mu\to e\gamma$ Right: $\tau\to e\gamma$ vs $\mu\to e\gamma$. On all panels, blue points correspond to $s_e=0.2$, while red points indicate $s_e>0.1$. Gray region is excluded by MEG, and dashed line indicates the reach of the future MEG upgrade.} 
\label{fig:lfv}
\end{figure*}

The plausibility of observing high-energy LFV can be described using ratios of branching ratios. For simplicity, we shall only consider the decay of $\tilde e_L$, which can be produced either directly, or from (wino-like) chargino decay. Given the values for the parameters favoured by neutrino data, we find:
\begin{eqnarray}
 \frac{\BR(\tilde e_L\to\mu\tilde\,\chi_0)}{\BR(\tilde e_L\to e\tilde\,\chi_0)} \approx s_e^2\lambda_L^2\epsilon^2\sim10^{-7}s_e^2~, \\
\frac{\BR(\tilde e_L\to\tau\tilde\,\chi_0)}{\BR(\tilde e_L\to e\tilde\,\chi_0)} \approx c_e^2\lambda_L^2\epsilon^2\sim10^{-7}c_e^2~,
\end{eqnarray}
where we have neglected lepton masses. Thus, the observation of LFV in selectron decays is unforeseeable in the near future.

The presence of light selectrons, along with the lack of significant mixing effects, means we would only expect electrons in final states when performing slepton or chargino searches at colliders. This fact could modify current LHC bounds on slepton or chargino masses, as such searches generally consider degenerate smuons and selectrons~\cite{CMS:2013dea,TheATLAScollaboration:2013hha}.

On the other hand, low-energy LFV takes into account radiative, leptonic and semileptonic LFV decays, such as $\mu\to e\gamma$, $\tau\to3e$ and $\tau\to\mu\eta$, as well as $\mu-e$ conversion in nuclei. In SUSY, it is well known that radiative LFV decays give the strongest bounds on slepton mixing (see~\cite{Arana-Catania:2013nha} for a recent review).


For the latter, given the size of the mixings and the strength of the different couplings, we find that $\mu\to e\gamma$ and $\tau\to\mu\gamma$ are dominated by the neutralino $RR$ contribution, while the main contribution to $\tau\to e\gamma$ is given by charginos with $LL$ mixing. For the first two, it is useful to compare the dominant contributions:
\begin{eqnarray}
\label{rat:megtmg} 
\frac{\BR(\mu\to e \gamma)}{\BR(\tau\to\mu\gamma)} &\approx&
 \left(\frac{m_\mu}{m_\tau}\right)^5\frac{\Gamma_\tau}{\Gamma_\mu}
\left|\frac{\mathcal{R}^R_{21}\mathcal{R}^{R*}_{L11}}
{\mathcal{R}^R_{31}\mathcal{R}^{R*}_{21}}\right|^2 \no\\
&\sim& 5.1\frac{1}{c_e^2\lambda_e^2\epsilon^2}y_\tau^2~.
\end{eqnarray}
This result implies that the branching ration of $\tau\to\mu\gamma$ is suppressed by a factor $\ord{10^{-7}}$ compared to that of $\mu\to e\gamma$. Considering the current bound on the latter by the MEG experiment~\cite{Adam:2013mnn}, this means that $\tau\to\mu\gamma$ shall be impossible to measure at Belle-II~\cite{Hayasaka:2013dsa}.

For $\tau\to e\gamma$, given the different chargino and neutralino masses and mixings, and taking into account the different loop functions, one cannot come up with a similar, reliable approximation.

In order to compare our results with those of~\cite{Blankenburg:2012nx}, we follow an identical scan: we fix $\tan\beta=10$, $\mu=600$~GeV, $M_2=500$~GeV, and set $M_1=0.5M_2$. The complete $6\times6$ slepton mass matrix and the $3\times3$ sneutrino mass matrix are diagonalised on the basis where $Y_e$ is diagonal. We take $m_\ell^2$ in the range $(200~{\rm GeV})^2-(1000~{\rm GeV})^2$ and $m^2_h$ between $5^2$ and $100^2$ times heavier. The $A_0$ parameter is assumed to be proportional to the heavy sfermion mass with a proportionality constant in the range $[-3,3]$.

In our numerical simulation, we include both chargino- and neutralino-mediated contributions to the three radiative LFV decays. Results are shown in Figure~\ref{fig:lfv}. In blue, we fix $s_e=s_d=0.2$, similar to what was done in~\cite{Blankenburg:2012nx}. In red, we vary $s_e>0.1$.

On the left panel of the Figure, we show the branching ratio of $\tau\to\mu\gamma$ versus that for $\mu\to e\gamma$. As expected, the former is very strongly suppressed. For the latter, we see that the current bounds of MEG already rule out some points, especially those for larger values of $s_e$. This implies a lower bound for the lightest selectron of about $700$~GeV. Nevertheless, there are still many points left, which can be almost fully probed by the future MEG upgrade~\cite{Baldini:2013ke}. This has the capacity of ruling out selectrons with masses up to $1.1$~TeV.

On the right panel, we show the correlation between the $\tau\to e\gamma$ and $\mu\to e\gamma$ branching ratios. We find their order of magnitude to be similar. Thus, as for $\tau\to\mu\gamma$, the branching ratio of $\tau\to e\gamma$ is expected to be too low to be measured at Belle-II.

With this, we confirm that this framework protects most flavoured processes in the lepton sector from SUSY contributions, to a sufficient degree. Without any doubts, the best way to probe our ansatz is through $\mu\to e\gamma$ decay, and through the detection of selectrons as the lightest sleptons at the LHC.

\section{Conclusions}
\label{sec:conc}

In this work we have explored an MFV-like ansatz based on $U(2)^5$ flavour symmetry, compatible with split-family SUSY. On its original version, this symmetry acted on the first two generation superfields, leaving the third generation as a singlet.

The main difference of this framework in comparison to the ones studied in~\cite{Barbieri:2011ci} and~\cite{Blankenburg:2012nx} is the inclusion of three $U(1)_f$ symmetries. While the $U(1)_d$ symmetry provides a justification for the difference between the top and bottom quark masses, the other two symmetries, acting on the lepton sector, provide a much stronger suppression. In fact, after ordering the charged lepton Yukawa matrix in terms of increasing eigenvalue, the $U(2)$ singlets become the first generation superfields, thus shifting the $U(2)^2$ symmetry towards the second and third generation.

The shifting of $U(2)^2$ allows the reproduction of the neutrino observables without any fine-tuning. In addition, the sfermion sector can have family splitting with no need of special cancellations. The consequence of this is a SUSY spectrum consisting at least of light stops, sbottoms and selectrons.

We find that the use of $U(2)^5$ symmetries protects flavoured processes very efficiently from SUSY contributions. This is done to the point of leaving $\mu\to e\gamma$ as the only flavour observable in the lepton sector that could be observed in the near future. In fact, the current bound by the MEG experiment already excludes selectrons of mass lower than $700$~GeV. If the future upgrade does not observe any signal, it would be able to rule out masses lower than $\sim 1100~{\rm GeV}$.

\section*{Acknowledgements}
I would like to thank Alberto Gago, Alvaro Ball\'on and especially Claudia Hagedorn for useful discussions, and Lorenzo Calibbi and Robert Ziegler for comments on the draft. I would also like to thank the University of Padua for the hospitality during his visit, and acknowledge support from the grants Generalitat Valenciana VALi+d, Spanish MINECO FPA 2011-23596 and the Generalitat Valenciana PROMETEO - 2008/004.

\appendix

\section{Normal Hierarchy}
\label{app:nh}

\begin{figure*}[tbp]
\begin{center}
\includegraphics[width=0.45\textwidth]{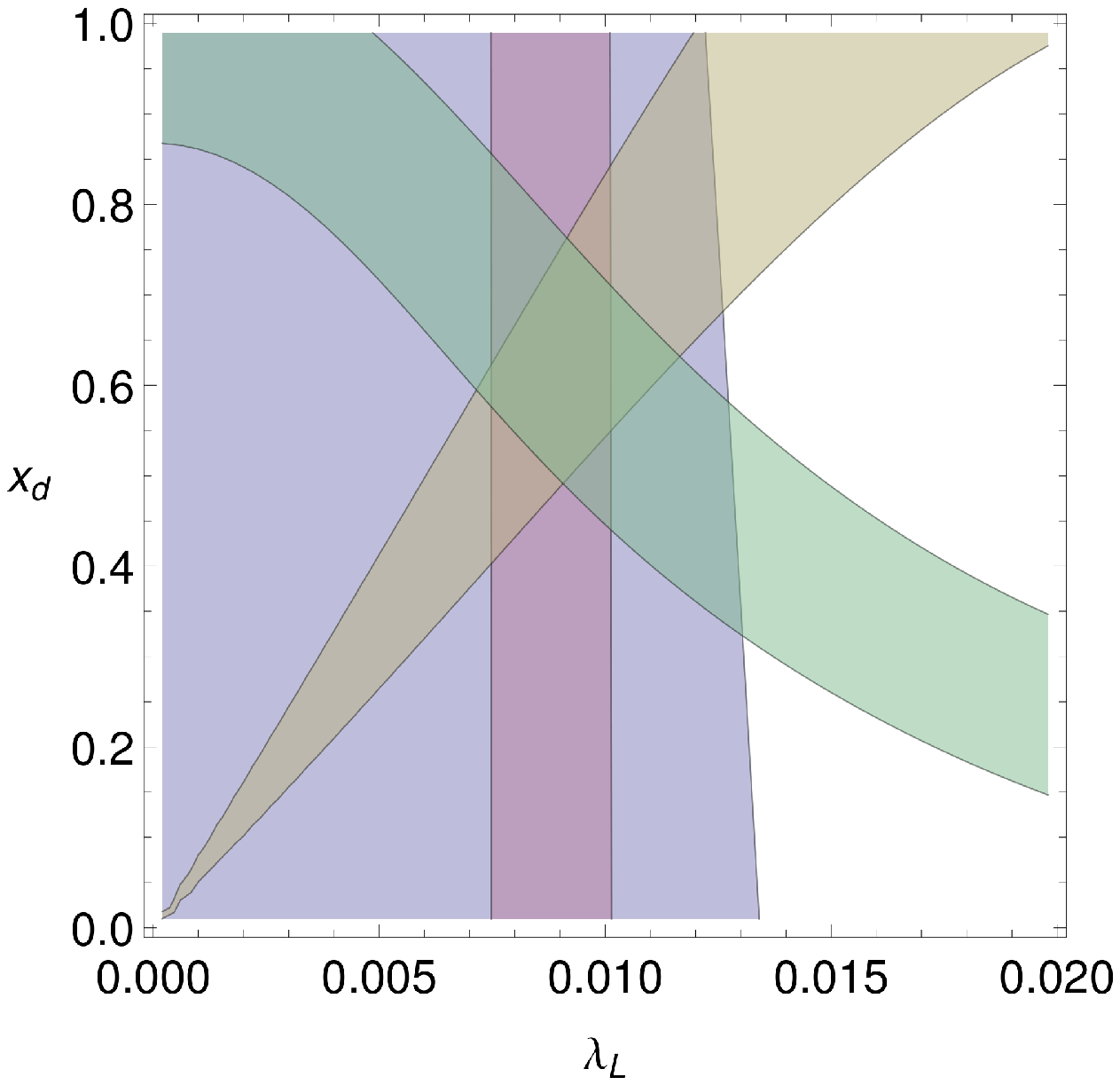} \quad
\includegraphics[width=0.45\textwidth]{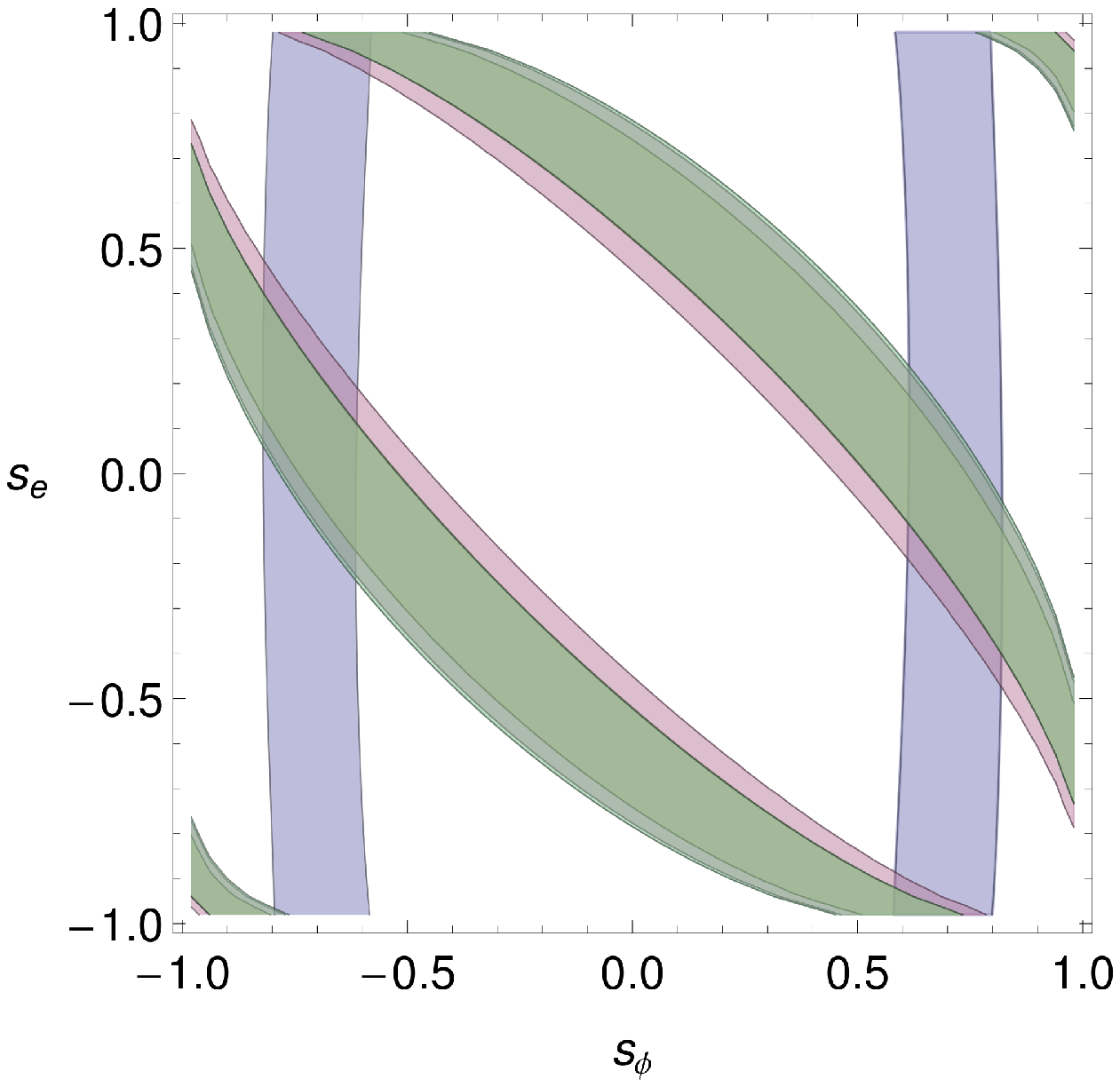}
\end{center}
\caption{Results for normal hierarchy. Left: Regions on the $x_d-\lambda_L$ plane satisfying $3\sigma$ constraints on oscillation parameters. Here we have $s_e=0$ and $s_\phi=0.63$. Right: The same on the $s_e-s_\phi$ plane. We set $x_d=0.6$ and $\lambda_L=9\times10^{-3}$. On all panels, blue, brown, purple and green regions satisfy $s^2_{23}$, $s^2_{12}$, $s^2_{13}$ and $\zeta^2$, respectively.} 
\label{fig:d1d2lambda}
\end{figure*}

The neutrino PMNS matrix can be obtained by diagonalising the neutrino mass matrix on the basis where $Y_e$ is diagonal. Thus, before addressing $m_\nu$, we shall first discuss the need to include information from the diagonalisation of $Y_e$.

Inspection of Eq.~(\ref{yuktexture}) shows that the main suppression parameters for the off-diagonal terms are $\lambda_e$ and $\epsilon$. For large enough $\lambda_e$ and low $\tan\beta$, one finds that the diagonalisation of $Y_e$ can significantly modify the muon Yukawa eigenvalue, leading to fine-tuning in order to correctly describe the muon mass.

To avoid this fine-tuning, one needs to impose upper bounds on $\lambda_e$. We find that, for $\tan\beta=10$, we need $\lambda_e\lesssim5\times10^{-2}$ if we want to keep the modification of the muon Yukawa lower than $1\%$. Moreover, for such values of $\lambda_e$, we find that the mixing angles diagonalising $Y_e\,Y_e^\dagger$ are at most $\ord{10^{-3}}$. Thus, we can ignore the contribution from the charged lepton sector when calculating the PMNS mixing angles.

We now turn to $m_\nu$. Given that current neutrino data provides only three mixing angles and two mass squared differences, we find this framework to have too many parameters in order to make any prediction on that sector. Thus, we use neutrino data as constraints on our parameters.

We find that taking the $\lambda_L\to0$, $d_1\to0$ limit reproduces neutrino data on the limit of Eq.~(\ref{eq-start.nh}), with the lightest neutrino mass also vanishing. If we keep $\lambda_L=0$, but take non-vanishing $d_1$, we find:
\begin{align}
 \zeta^2=\frac{d_1^2}{d_2^2}~, & & s_{23}^2=s_\phi^2~,
\end{align}
with all other mixing angles equal to zero. The observed $\zeta^2$ indicates the need of a hierarchy between $d_1$ and $d_2$.

When taking $\lambda_L\neq0$, we generate the other mixing angles, and find that the above relations are disturbed. Nevertheless, this contribution is small, and $\zeta^2$ can obtain its correct value by adjusting $d_1=x_d\,\zeta_{\rm exp}\,d_2$, with $x_d$ of $\ord{1}$.

On Figure~\ref{fig:d1d2lambda}, we show the regions that satisfy the $3\sigma$ constraints on neutrino oscillation parameters on the $x_d-\lambda_L$ and $s_e-s_\phi$ planes. Other parameters are fixed, and no phases are taken into account. On the left panel, we see that $s^2_{12}$ and $\zeta^2$ show opposite behaviours on the $x_d-\lambda_L$ plane, such that they intersect only on a specific region. This region can be made to coincide with the one reproducing the correct $s^2_{13}$. The Figure suggests taking $\lambda_L=10^{-2}$, with deviations being taken into account by the $\ord{1}$ parameters appearing in Eq.~(\ref{majomat}).

By setting $\lambda_L$ at this value, and adjusting $d_1$ to reproduce the observed hierarchy, one can do a more detailed analysis of $s^2_{13}$ and $s^2_{12}$, taking into account the variation of all parameters. For instance, $s^2_{13}$ depends on the framework parameters through:
\begin{equation}
 \label{eq:s13analytic}
 s_{13}\approx\frac{r_2}{d_2}\frac{\lambda_L}{\epsilon}|c_ec_\phi-e^{i\omega_3}s_es_\phi|~.
 \end{equation}
This explains the correlation between $s_e$ and $s_\phi$, seen on the right panel of Figure~\ref{fig:d1d2lambda}. All parameters, with the exception of $s^2_{23}$, shall present a similar behaviour with respect to these angles.

In principle, Equation~(\ref{eq:s13analytic}) can be used to determine which values of $\lambda_L/d_2$ reproduce best the reactor angle. However, we find that for $d_2$ larger than $\ord{1}$, corresponding to $\lambda_L>10^{-2}$, the intersection of regions for $s^2_{12}$ and $\zeta^2$ become much smaller. Thus, we shall set $d_2$ as an $\ord{1}$ parameter, which justifies our choice of $\lambda_L=10^{-2}$. This means that the elements of $\Delta_L$ can be at most of order $\epsilon^2$.

The equation for $s^2_{12}$ is quite complicated. On the CP-conserving scenario, one finds:
\begin{multline}
\label{eq:s12analytic}
\frac{1}{2}\tan2\theta_{12}\approx\frac{r_2}{d_2}\frac{\lambda_L}{\epsilon}\frac{(s_e c_\phi+c_e s_\phi)}{d_1^2\epsilon^2-r_1^2\lambda_L^2/\epsilon^2}\cdot \\
 \left(d_1\,d_2\,\epsilon^2+r_1d_2\lambda_L^2-r_2^2\left(c_ec_\phi-s_es_\phi\right)^2\right)~,
\end{multline}
with all terms being relevant for the determination of the solar mixing angle.

To summarise, the charged lepton and neutrino sector demand the following constraints on the framework parameters:
\begin{align}
 \lambda_L = 10^{-2}~, & & \lambda_e = y_e\,10^2~, \\
 d_1=x_d\,\zeta_{\rm exp}d_2~, & & s^2_\phi\in\left[0.36,\, 0.67\right]~. 
\end{align}
The rest of the parameters must have values such that $s^2_{12}$ and $s^2_{13}$, determined through Eqs.~(\ref{eq:s12analytic}) and~(\ref{eq:s13analytic}), follow the experimental observations.

\section{Inverted Hierarchy}
\label{app:ih}


For the inverted hierarchy, if one takes the $s^2_{13}\to0$ and $\Delta m^2_{\rm sol}\to0$ limit, an equation similar to Eq.~(\ref{eq-start.nh}) is obtained:
\begin{multline}\label{eq-start.ih}
(M_\nu^2)_{\rm i.h.}\rightarrow m_{\rm light}^2 \cdot \left(\begin{array}{ccc}
1 & 0 & 0\\0 & 1 & 0\\0 & 0 & 1 
\end{array}\right)\\ + \Delta m^2_{\rm atm} \left(
\begin{array}{ccc}
1 & 0 & 0\\0 & c_{23}^2 & -s_{23}c_{23}\\0 & -s_{23}c_{23} & s_{23}^2 
\end{array}\right)~.
\end{multline}
To reach this limit, we require $d_2\to\lambda_L^2/\epsilon^2$, $d_1\to0$ and $\epsilon\to0$. Although this limit imposes only one more condition on our parameters, this new condition relates $d_2$ and $\lambda_L$, which are associated to different symmetries of the framework. This suggests that the inverted hierarchy might require some tuning.

To generate $\Delta m^2_{\rm sol}$, we need to take deviations of $d_2$, i.e.\ we set $d_2=x_\lambda \lambda_L^2/\epsilon^2$. For instance, if the value of $x_\lambda>1$, we get:
\begin{equation}
 |\zeta^2|=\frac{x_\lambda^2-1}{x_\lambda^2}~.
\end{equation}
Thus, $x_\lambda$ cannot be much more different than unity.

Variations of $d_1$ do not affect much our results. We set $d_1=y_\lambda\lambda_L^2/\epsilon^2$, where we assume $y_\lambda$ is small. For $x_\lambda>1$, this implies:
\begin{equation}
 |\zeta^2|=\frac{x^2_\lambda-1}{x^2_\lambda-y^2_\lambda}~.
\end{equation}
We see that $y_\lambda$ does not have to be vanishingly small, and we have checked that values as large as $10^{-1}$ do not disturb numerical results. Still, $x_\lambda$ must remain close to unity, meaning that non-vanishing $d_1$ does not make the situation any more flexible.

As one increases $\epsilon$ from zero, one generates non-zero $s^2_{12}$ and $s^2_{13}$. Moreover, for very large $\epsilon$ ($\sim\lambda^2_{\rm CKM}$), its contributions spoil the value of $\zeta^2$. Thus, similarly to the normal hierarchy, we need to modify our previous equations. In the following, we shall explore the following relation:
\begin{equation}
 d_2=x'_\lambda\frac{\lambda_L^2}{\epsilon^2}\sqrt{\frac{1-\zeta^2_{\rm exp}y_\lambda^2}{1-\zeta^2_{\rm exp}}}~,
\end{equation}
which corresponds to our previous limit for $x_\lambda>1$ when $x'_\lambda=1$.

\begin{figure*}[tbp]
\begin{center}
\includegraphics[width=0.45\textwidth]{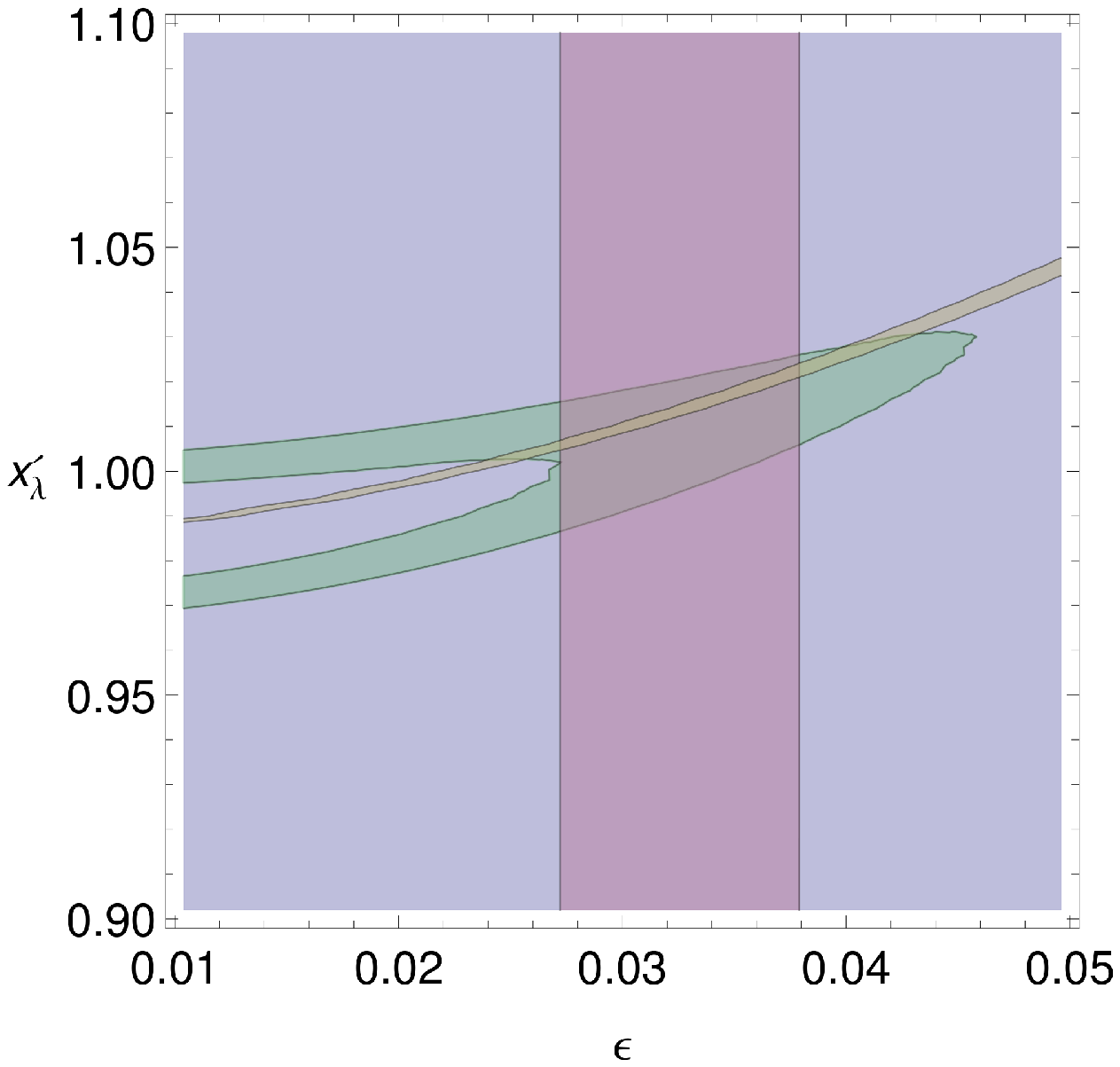} \quad
\includegraphics[width=0.45\textwidth]{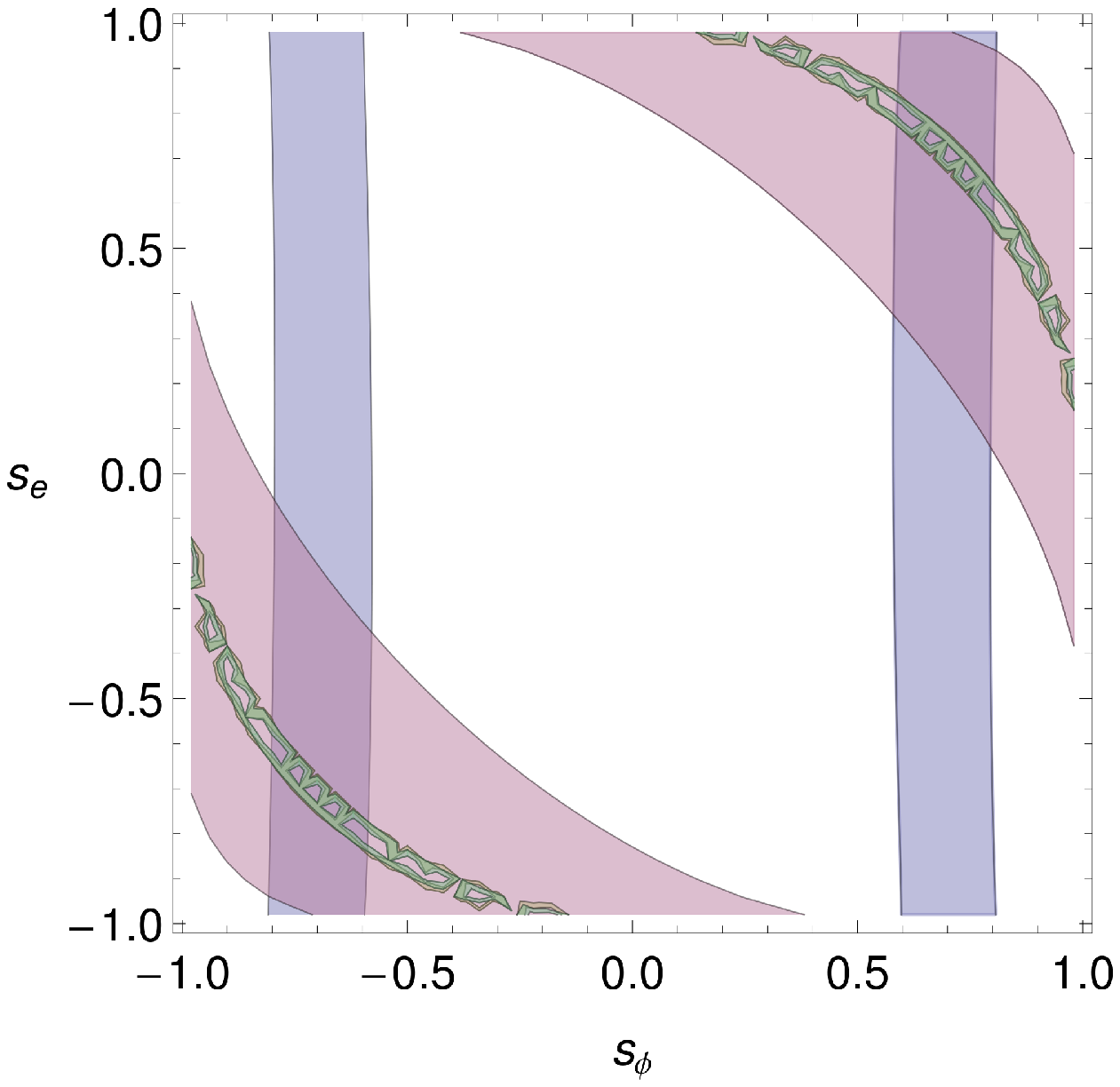}
\end{center}
\caption{Results for inverted hierarchy. Left: Regions on the $x'_\lambda-\epsilon$ plane satisfying $3\sigma$ constraints on oscillation parameters. Here we have $s_e=0.73$ and $s_\phi=0.65$. Right: The same on the $s_e-s_\phi$ plane. We set $x_\lambda=1.013$ and $\epsilon=3.25\times10^{-2}$. On all panels, blue, brown, purple and green regions satisfy $s^2_{23}$, $s^2_{12}$, $s^2_{13}$ and $\zeta^2$, respectively.} 
\label{fig:d1d2ih}
\end{figure*}
For our numerical analysis, we set $\lambda_L=\lambda_{\rm CKM}$, $y_\lambda=0$, and no CP phases. We show results in Figure~\ref{fig:d1d2ih}. On the left panel, we fix $s_\phi=0.65$ and $s_e=0.73$, and plot the $3\sigma$ confidence intervals on the $x'_\lambda-\epsilon$ plane. We find a small region that does satisfy all parameters. Nevertheless, its area is very small, as $s^2_{12}$ has a very small acceptable region.

The fine-tuning of the inverted hierarchy is also made evident on the right panel of the Figure, where we show the $s_e-s_\phi$ parameter space, for $x_\lambda=1.013$ and $\epsilon=3.25\times10^{-2}$. Clearly, both $s^2_{12}$ and $\zeta^2$ require very specific values of $s_e$, $s_\phi$ and $x'_\lambda$ to be properly reproduced. In fact, throughout the parameter space, the values of both oscillation parameters are generally much larger than the ones required, which makes this particular scenario very unattractive.

On the previous analysis, we have taken $\lambda_L=\lambda_{\rm CKM}$. Smaller values of $\lambda_L$ require smaller values of $\epsilon$, meaning that the inverted hierarchy does not allow us to follow the assumption of $\epsilon\sim\lambda_{\rm CKM}^2$. This is not a problem, but does not bring any improvement in the analysis.

One could also be tempted to use $\lambda_L=1$ by removing the $U(1)_L$ symmetry. This would require the $\Delta_L$ spurion not to be suppressed, which would bring issues with the perturbative expansion in the slepton sector.

To conclude, although the inverted hierarchy can provide regions in the parameter space that reproduce the $3\sigma$ observed values, these regions seem strongly fine-tuned. Apparently, we require two tunings, one between $\lambda_L$ and $d_2$, and another one between $s_\phi$ and $s_e$. The former tuning is the most disturbing, as it relates two parameters that have no symmetry connection between them.


\begin{thebibliography}{99}

\bibitem{manycites2}
  N.~Craig,
  arXiv:1309.0528 [hep-ph]~;
  J.~A.~Evans, Y.~Kats, D.~Shih and M.~J.~Strassler,
  arXiv:1310.5758 [hep-ph].

\bibitem{Papucci:2011wy}
  M.~Papucci, J.~T.~Ruderman and A.~Weiler,
  JHEP {\bf 1209} (2012) 035
  [arXiv:1110.6926 [hep-ph]].
  
\bibitem{Antusch:2012gv}
  S.~Antusch, L.~Calibbi, V.~Maurer, M.~Monaco and M.~Spinrath,
  JHEP {\bf 01} (2013) 187
   [JHEP {\bf 1301} (2013) 187]
  [arXiv:1207.7236].
  
\bibitem{Feng:1999mn}
  J.~L.~Feng, K.~T.~Matchev and T.~Moroi,
  Phys.\ Rev.\ Lett.\  {\bf 84} (2000) 2322
  [hep-ph/9908309].
  
\bibitem{TheATLAScollaboration:2013fha}
  The ATLAS collaboration,
  ATLAS-CONF-2013-047.
  
\bibitem{CMS:2013gea}
  CMS Collaboration [CMS Collaboration],
  CMS-PAS-SUS-13-012.
  
\bibitem{manycites1}  
  M.~Dine, R.~G.~Leigh and A.~Kagan,
  Phys.\ Rev.\ D {\bf 48} (1993) 4269
  [hep-ph/9304299]~;
  S.~Dimopoulos and G.~F.~Giudice,
  Phys.\ Lett.\ B {\bf 357} (1995) 573
  [hep-ph/9507282]~;
  A.~G.~Cohen, D.~B.~Kaplan and A.~E.~Nelson,
  Phys.\ Lett.\ B {\bf 388} (1996) 588
  [hep-ph/9607394]~;
  G.~F.~Giudice, M.~Nardecchia and A.~Romanino,
  Nucl.\ Phys.\ B {\bf 813} (2009) 156
  [arXiv:0812.3610 [hep-ph]]~;
  R.~Barbieri, E.~Bertuzzo, M.~Farina, P.~Lodone and D.~Pappadopulo,
  JHEP {\bf 1008} (2010) 024
  [arXiv:1004.2256 [hep-ph]]~;
  M.~Badziak, E.~Dudas, M.~Olechowski and S.~Pokorski,
  JHEP {\bf 1207} (2012) 155
  [arXiv:1205.1675 [hep-ph]].
  
\bibitem{D'Ambrosio:2002ex}
  G.~D'Ambrosio, G.~F.~Giudice, G.~Isidori and A.~Strumia,
  Nucl.\ Phys.\ B {\bf 645} (2002) 155
  [hep-ph/0207036].
  
\bibitem{Barbieri:2011ci}
  R.~Barbieri, G.~Isidori, J.~Jones-Perez, P.~Lodone and D.~M.~Straub,
  Eur.\ Phys.\ J.\ C {\bf 71} (2011) 1725
  [arXiv:1105.2296 [hep-ph]].

\bibitem{Barbieri:2011fc}
  R.~Barbieri, P.~Campli, G.~Isidori, F.~Sala and D.~M.~Straub,
  Eur.\ Phys.\ J.\ C {\bf 71} (2011) 1812
  [arXiv:1108.5125 [hep-ph]].
  
\bibitem{Barbieri:2012uh}
  R.~Barbieri, D.~Buttazzo, F.~Sala and D.~M.~Straub,
  JHEP {\bf 1207} (2012) 181
  [arXiv:1203.4218 [hep-ph]].
  
\bibitem{Buras:2012sd}
  A.~J.~Buras and J.~Girrbach,
  JHEP {\bf 1301} (2013) 007
  [arXiv:1206.3878 [hep-ph]].
  
\bibitem{Blankenburg:2012ah}
  G.~Blankenburg and J.~Jones-Perez,
  Eur.\ Phys.\ J.\ C {\bf 73} (2013) 2289
  [arXiv:1210.4561 [hep-ph]].
  
\bibitem{Blankenburg:2012nx}
  G.~Blankenburg, G.~Isidori and J.~Jones-Perez,
  Eur.\ Phys.\ J.\ C {\bf 72} (2012) 2126
  [arXiv:1204.0688 [hep-ph]].

\bibitem{GonzalezGarcia:2012sz}
  M.~C.~Gonzalez-Garcia, M.~Maltoni, J.~Salvado and T.~Schwetz,
  JHEP {\bf 1212} (2012) 123
  [arXiv:1209.3023 [hep-ph]],
  \newblock updated results available at {\tt http://www.nu-fit.org}.
  
\bibitem{Cirigliano:2005ck}
  V.~Cirigliano, B.~Grinstein, G.~Isidori and M.~B.~Wise,
  Nucl.\ Phys.\ B {\bf 728} (2005) 121
  [hep-ph/0507001].
  
\bibitem{Schwingenheuer:2012zs}
  B.~Schwingenheuer,
  Annalen Phys.\  {\bf 525} (2013) 269
  [arXiv:1210.7432 [hep-ex]].
  
\bibitem{Jimenez:2010ev}
  R.~Jimenez, T.~Kitching, C.~Pena-Garay and L.~Verde,
  JCAP {\bf 1005} (2010) 035
  [arXiv:1003.5918 [astro-ph.CO]].
  
\bibitem{Delgado:2011kr}
  A.~Delgado and M.~Quiros,
  Phys.\ Rev.\ D {\bf 85} (2012) 015001
  [arXiv:1111.0528 [hep-ph]].

\bibitem{Dudas:2013pja}
  E.~Dudas, G.~von Gersdorff, S.~Pokorski and R.~Ziegler,
  arXiv:1308.1090 [hep-ph].
  
\bibitem{Abada:2010kj}
  A.~Abada, A.~J.~R.~Figueiredo, J.~C.~Romao and A.~M.~Teixeira,
  JHEP {\bf 1010} (2010) 104
  [arXiv:1007.4833 [hep-ph]].
  
\bibitem{Abada:2012re}
  A.~Abada, A.~J.~R.~Figueiredo, J.~C.~Romao and A.~M.~Teixeira,
  JHEP {\bf 1208} (2012) 138
  [arXiv:1206.2306 [hep-ph]].
  
\bibitem{CMS:2013dea}
  CMS Collaboration [CMS Collaboration],
  CMS-PAS-SUS-13-006.
  
\bibitem{TheATLAScollaboration:2013hha}
  The ATLAS collaboration,
  ATLAS-CONF-2013-049.
  
  
\bibitem{Arana-Catania:2013nha}
  M.~Arana-Catania, S.~Heinemeyer and M.~J.~Herrero,
  Phys.\ Rev.\ D {\bf 88} (2013) 015026
  [arXiv:1304.2783 [hep-ph]].
 
\bibitem{Adam:2013mnn}
  J.~Adam {\it et al.}  [MEG Collaboration],
  arXiv:1303.0754 [hep-ex].
 
\bibitem{Hayasaka:2013dsa}
  K.~Hayasaka [Belle and Belle II Collaboration],
  J.\ Phys.\ Conf.\ Ser.\  {\bf 408} (2013) 012069.
   
\bibitem{Baldini:2013ke}
  A.~M.~Baldini, F.~Cei, C.~Cerri, S.~Dussoni, L.~Galli, M.~Grassi, D.~Nicolo and F.~Raffaelli {\it et al.},
  arXiv:1301.7225 [physics.ins-det].
  
  
\end{thebibliography}
\end{document}